# FROM MMU TO MPU: ADAPTATION OF THE PIP KERNEL TO CONSTRAINED DEVICES

Nicolas Dejon[1][2], Chrystel Gaber[1] and Gilles Grimaud[2]

[1]Orange Labs, Châtillon, France
nicolas.dejon@orange.com
chrystel.gaber@orange.com

[2]Univ. Lille, CNRS, Centrale Lille, UMR 9189 CRIStAL - Centre de Recherche en Informatique Signal et Automatique de Lille, F-59000 Lille, France
gilles.grimaud@univ-lille.fr

## ABSTRACT

*This article presents a hardware-based memory isolation solution for constrained devices. Existing solutions target high-end embedded systems (typically ARM Cortex-A with a Memory Management Unit, MMU) such as seL4 or Pip (formally verified kernels) or target low-end devices such as ACES, MINION, TrustLite, EwoK but with limited flexibility by proposing a single level of isolation. Our approach consists in adapting Pip to inherit its flexibility (multiple levels of isolation) but using the Memory Protection Unit (MPU) instead of the MMU since the MPU is commonly available on constrained embedded systems (typically ARMv7 Cortex-M4 or ARMv8 Cortex-M33 and similar devices). This paper describes our design of Pip-MPU (Pip's variant based on the MPU) and the rationale behind our choices. We validate our proposal with an implementation on an nRF52840 development kit and we perform various evaluations such as memory footprint, CPU cycles and energy consumption. We demonstrate that although our prototyped Pip-MPU causes a 16% overhead on both performance and energy consumption, it can reduce the attack surface of the accessible application memory from 100% down to 2% and the privileged operations by 99%. Pip-MPU takes less than 10 kB of Flash (6 kB for its core components) and 550 B of RAM.*

## KEYWORDS

*constrained devices, MPU, memory isolation, Pip, OS kernel, secure systems*

## 1. INTRODUCTION

Given the growing ubiquity of low-end devices (sensors, actuators) that can be managed remotely through the Internet, preventing remote cyberattacks leveraging these devices requires isolating sensible functionalities from untrusted ones. High-end devices, like servers and traditional computers, already propose strong security mechanisms such as Pip [1], seL4 [2] or mC2/CertiKOS [3] which all ensure memory isolation between memory spaces by the use of the Memory Management Unit (MMU).

However, constrained devices is a category of devices outlining limited resources compared to high-end devices in terms of memory, computing power and energy supply. Class 2 [4] low-end microcontrollers are constrained devices enough capable of supporting full protocol stacks so to easily connect to the Internet Of Things (IoT), while being limited in memory (>50 KB RAM and >250 KB Flash). For memory protection, they might be equipped only with a Memory Protection Unit (MPU), like the majority of the boards based on the ARM Cortex-M processor family [5], which do not offer memory virtualisation. Therefore, existing formally verified isolation kernels

[1-3] cannot be used for these targets and existing isolation solutions such as ACES, MINION, TrustLite, Ewok [6-9] are limited because they offer only one level of isolation.

The motivation of this work is to make constrained devices more secure and more flexible, given the ubiquity of these devices and the emerging complex IoT applications. We propose to achieve flexible memory isolation for constrained devices with an MPU by adapting Pip's MMU-based memory isolation to be MPU-based and by leveraging Dejon et al.'s framework [10]. Indeed, this framework proposes to use the memory access permissions on memory blocks provided by the MPU in order to create multiple levels of isolation. To achieve this objective, we investigate the following questions: i) how can the framework be specialised with Pip's security requirements? ii) can Pip's flexibility be adapted to constrained objects with MPU? iii) what are the costs of porting an existing system on this MPU-based solution?

To the best of our knowledge, this is the first time a transposition of this nature (MMU to MPU without loss of features) is realised.

Our main contributions are as follows:
- We capture and define Pip's requirements that are landmarks to our adaptation.
- We specialise the framework presented in [10] to match the aforementioned requirements. We also conduct a preliminary study of compatibility between Pip and the framework.
- We implement the specialisation on an ARMv7 Cortex-M processor-based device with MPU, calling it Pip-MPU. It is the first implementation of Dejon et al.'s framework and therefore the first system proposing nested compartmentalisation for constrained devices.
- We thoroughly evaluate our Pip-MPU prototype in terms of CPU cycles, initialisation time, memory footprint and energy consumption overhead. The analysis also covers security metrics such as accessible memory areas and privileged cycles.

Formal verification of the security properties, paired with Pip, is an ongoing work not covered in this paper.

The rest of the paper is constructed as explained in the following. We discuss related work in Section 2. Then, a preliminary background is given in Section 3, gathering a brief overview of Pip's architecture and requirements, as well as a succinct presentation of the MPU. In Section 4, we present Pip-MPU's requirements that include Pip's requirements plus some requirements specific to constrained devices. In Section 5, we verify which requirements are already satisfied by the use of the nested compartmentalisation framework [10]. We then derive and specialise this framework in the light of Pip's system calls and metadata structures to fulfil the security requirements. We discuss the design choices and end up with a full implementation of Pip-MPU. In Section 6, we evaluate the implementation on an ARM Cortex-M4 (ARMv7-M architecture) device. ARM Cortex-M devices have widespread use among IoT (Internet-of-Things) vendors. We perform the evaluation on performance and security metrics to assess the solution's industrial viability and the fulfilment of Pip-MPU's requirements.

## 2. RELATED WORK

The research community invested many efforts in MPU-based security architectures (ACES, MINION, TrustLite, EwoK, TockOS, OPEC [6-9, 11-12]). Unfortunately, they are not suitable for Pip's design as they mostly have a security policy that is fixed at design time (e.g. before runtime) while few systems like TockOS offer dynamic application loading during runtime. Furthermore, all consider flat memory isolation compared to the hierarchical partitioning design of Pip that Pip-MPU inherits. Some systems also compromise the compartmentalisation like ACES because of the mentioned MPU limitations whereas Pip-MPU can deal with any number of partitions without loss of isolation. In addition to that, Pip-MPU just needs two reserved MPU regions while all other mentioned systems further limit the user configurable MPU regions or assign specific memory types to them (code, data, peripherals…). In addition to that, Pip-MPU is not tied to a specific architecture like the systems above because the nested compartmentalisation framework is compatible both with the ARMv7 and ARMv8 architectures. More than that, the ARMv8 architecture releases the reserved MPU regions constraint because of the MPU region alignment constraints that just apply to the ARMv7 architecture.

General-purpose systems usually have well-established security mechanisms. Efforts towards formally verified systems like Pip resulted notably in high-assurance systems like seL4 [2] and mC2/CertiKOS [3]. However, these systems target high-performance computers and are tied to their hardware platform, not suitable for low-end devices because of an absence of technology or economic incentives. In an IoT ecosystem that is dynamic and demanding, data and applications from low-end devices must be protected in order to transmit correct information to decision makers. Pip-MPU is also meant to be formally verified following Pip's proof methodology.

Memory isolation techniques for constrained devices are manyfold. Previously discussed MPU-based systems are hardware-rooted but there are hybrid approaches extending the list like TyTAN [13] based on Trustlite, SMART [14], Sancus [15], CheriRTOS [16]. However, they all modify the hardware in a way, for example by extending the CPU instructions or enhancing memory bus access logic. While they show reasonable performance for the embedded systems use cases, the required hardware customisation may be too expensive for low-end devices. Pip-MPU does not modify the hardware nor extend ARM's ISA. By using widely available hardware, Pip-MPU can be used for COTS systems thus keeping production costs low. This way, Pip-MPU keeps its software layers minimal, exposing a small TCB and reducing the attack surface. There also exists software-only memory isolation techniques as illustrated in the Security microvisor [17]. However, the latter also suffers from the unique segregation between a secure world and a non-secure world while Pip-MPU offers multiple isolation levels. PISTIS [18] is another software-only solution for constrained devices deprived of MPU that adds an onboard application verifier and loader. Other systems (i.e. MINION, ACES) also need additional firmware analysis, either offline or when an application is loaded. Pip-MPU avoids the struggle of application verification since the partition is free to evolve as it wishes within the MPU harness set up by Pip-MPU.

Other hardware modules than the MPU exist and are sometimes used to set up enclaves for memory isolation like the ARM TrustZone [19] for ARM architectures or Intel SGX [20] and Memory Protection Keys (MPK) [21] for high-end Intel machines. Nevertheless, they stay limited in the number of protected domains compared to what is proposed with Pip-MPU. However, Pip-MPU can be complementary to some enclaves, for example implemented with TrustZone-enabled devices as MPUs might be present in both secure and non-secure worlds.

## 3. BACKGROUND

Pip is a Trusted Computing Base (TCB) that provides only data isolation and control flow handling features. Therefore, it is either used by single-thread and multi-tasking bare-metal applications or by an OS that provides additional properties such as scheduling, Inter-process Communication (IPC) and drivers. Pip's API is comprised of a dozen system calls, covering memory management and context switching.

### 3.1. Pip partitioning model

Pip's memory management is based on a hierarchical partitioning model.

The main principle is that a *partition* (an execution unit) can create one or several subpartitions that in turn can create subpartitions. This creates a partition tree as can be seen in Figure 1, rooted in a special partition called the *root partition*. The root partition is the only partition existing at system initialisation. The other partitions are dynamically created by the user during the system's lifetime.

Pip's security goal is spatial memory isolation which is set up by partitioning. Pip protects the data confidentiality and integrity of all partitions by memory isolation. No partitions should access a particular partition's private data, except the memory shared with descendants or ancestors in the partition tree. Furthermore, Pip registers the partition tree in its metadata structures. These structures should be protected and remain isolated from any partition, otherwise partitions could grant themselves permissions on memory they don't own. Pip's code integrity should also be ensured for Pip's proven properties to hold.

Pip enforces 3 security properties at any time providing rules for data isolation and sharing:
- **Kernel isolation** Strict memory isolation between the kernel (code, data and metadata structures) and the partitions;
- **Vertical sharing** Any memory owned by a partition is shared with its unique parent;
- **Horizontal isolation** Strict memory isolation between sibling partitions or partitions branched from an ancestor.

This means memory owned by the parent can only be attributed to a single child (no shared memory in siblings).

These properties are represented in Figure 1.

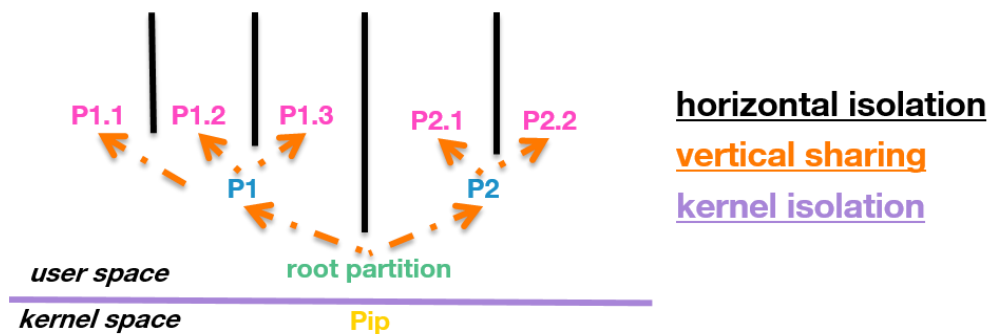

Figure 1. Pip's partitioning scheme.

## 3.2. Pip architecture

Pip is built on a stack of layers represented in Figure 2: partitions, LibPip, pipcore, the Memory Abstraction Layer (MAL) and the hardware platform. The layers are split into a kernel space (where Pip lies) and a user space (also called userland).

In userland lie the partitions. Partitions can directly call Pip's services or use LibPip, the library dedicated to do the system calls. LibPip also lies in userland and is made of two sublayers, a low-level API and a high-level API. The lower level crafts the requests to the system calls, setting the parameters in the correct registers and making the system call. While this is enough to make raw use of Pip's services, LibPip's higher level is intended to facilitate the user's interactions with Pip. For example, it could be a dedicated function to set up and launch a child partition. This level is context dependent and uses LibPip's lower level.

In kernel space lie Pip's core services, referenced as pipcore. They are the set of services exposed by Pip to the partitions. Pipcore is Pip's main component including the algorithms configuring the hardware. Because of this sensitive nature, pipcore provides proof of memory isolation preservation on these system calls. Consequently, pipcore is directly written into the Coq Proof Assistant [22] where the proofs can also be conducted. An inner custom tool then translates this code into C that is later compiled with the other software layers altogether.

Pipcore delegates all the reads and writes to the lower software layer called the Memory Abstraction Layer (MAL). This layer directly interacts with the machine and is made of simple memory operations. It hides the memory interaction details to pipcore which ensures pipcore's portability. The MAL is part of a bigger lower layer of trusted components which contains additional procedures and handlers in C and ASM. This latter layer encompasses Pip's initialisation sequence (root partition launch), the board's boot procedure, the exception handlers and other implementation dependant routines. This boot procedure should be adapted accordingly to the hardware platform and must be privileged to access system peripherals.

Finally, the hardware platform encompasses the MMU. Pip-MPU carries on Pip's method to build a security kernel fitted to conduct formal proofs. As such, we keep the same architecture for Pip-MPU's design, with the hardware platform based on the MPU.

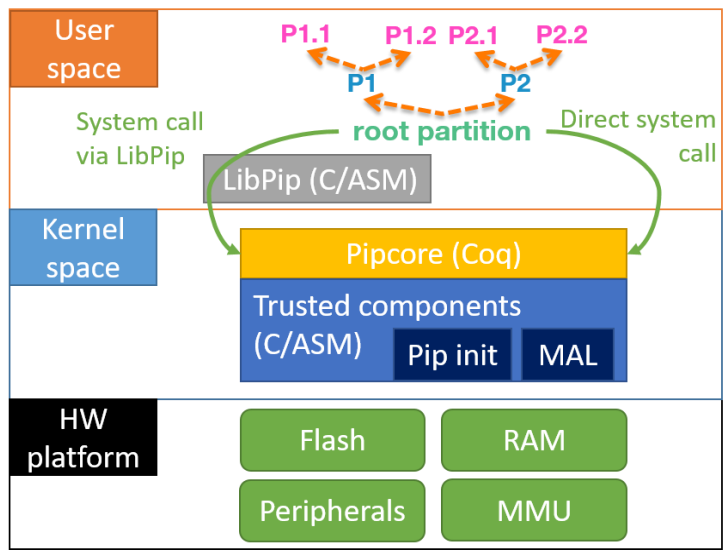

Figure 2. Pip's architecture. Partitions can directly invoke Pip system calls or pass via LibPip.

### 3.3. The Memory Protection Unit (MPU)

The MPU ensures hardware-based memory protection similar to the MMU, but does not virtualize the memory. As a consequence, the MPU organizes the space in MPU regions, i.e. continuous ranges of memory addresses of variable size, whereas the MMU organises the space by memory pages, usually of fixed size. Since the memory is smaller for devices with MPU, typically 8-16 MPU regions can be configured and protected at the same time. The MPU's configuration is stored in CPU registers while the MMU manages page tables stored in the main memory. The MPU regions play the same role as MMU pages and can be hardware protected with associated access control rights. As with MMU pages, illegal access ends up in a memory fault.

We summarize the key differences between MMU and MPU in Table 1.

The highlighted differences prevent us from directly transposing from an MMU-based system to a system based on an MPU. The limited number of MPU regions, designed accordingly to constrained devices' requirements, doesn't scale with the millions of pages protected by an MMU. Furthermore, they are configured and they operate so differently that the configuration software should be entirely redesigned. As a consequence, Pip can't be used on devices without the MMU hardware like our targeted constrained devices. This implies a radical change in pipcore and the MAL which are tightly coupled to the hardware platform.

Table 1. MMU versus MPU.

| Attributes | MMU | MPU |
|---|---|---|
| Virtual memory | Yes | No |
| Configuration mode | Privileged | Privileged |
| Memory region unit | Page | MPU region |
| Number of memory region unit | Millions | 8-16 |
| Access control (RWX) | Yes | Yes |
| Configuration storage | Main memory | Registers |
| Device memory size | MB-GB | kB |
| Device frequency | GHz | MHz |

## 4. PIP-MPU'S REQUIREMENTS

This section defines the requirements that Pip-MPU must satisfy. We classified the requirements into four categories: security requirements, performance requirements, functional requirements and hardware requirements. Some requirements are directly inherited from Pip while the others are required to target resource-constrained low-end devices.

### 4.1. Pip's fundamental requirements

Pip-MPU inherits all Pip's requirements, outside the ones tied to the MMU. Hence, we first state and classify the set of Pip's fundamental requirements.

- **SecReq1**: Pip's security properties. Pip's security properties described in Section 3.1 shall be ensured.
- **SecReq2**: Hardware-based memory protection. Any illegal access shall be blocked and identified by the hardware-based memory protection components. Only the kernel space has sufficient privileges to configure them.
- **SecReq3**: Minimal software size. Pip's code must be minimal in size in order to be formally verified, to reduce the likelihood of vulnerabilities, and to ease the maintenance of the code base.
- **SecReq4**: Limited access permissions updates. Pip shall ensure that only a parent partition can manage block access permissions (read, write, execution), that might be changed during the partition's lifetime. Pip shall ensure that a partition cannot increase the rights set up by the parent partition, on itself or one of its children.
- **FuncReq1**: Flexible partitions. The partition tree shall be determined at runtime. Any partition can create and isolate a subspace of its own.
- **PerfReq1**: Reasonable performance overhead. Pip shall maintain the performance requirements existing before the port to Pip in order to address real-world scenarios. This includes a fast startup sequence (fast cold start) that should not significantly impact the bootstrapping routine.

### 4.2 Specific Pip-MPU requirements

In a second step, we define additional performance and hardware requirements that stem from the constrained nature of the targeted devices. Indeed, Pip-MPU targets devices without MMU and is challenged by their constrained resources.

- **HWReq1**: MPU-based memory protection. Pip-MPU shall specifically use the MPU as hardware memory protection. As the MPU is only present in low-end devices, the corollary is that Pip-MPU only targets this class of devices.

- **HWReq2**: No hardware modifications. Pip-MPU shall use hardware components present in Commercial Off-The-Shelf (COTS) systems, without any hardware modifications. This is to ease its adoption and reduce development time.
- **PerfReq2**: Bounded execution time. Pip-MPU's algorithm complexities and implemented code shall be compatible with real-time constraints. Indeed, many low-end device scenarios have such constraints.
- **PerfReq3**: Low memory consumption. Pip-MPU shall let enough space for real-world scenarios to fit in a Pip-based system. Pip-MPU's security overlay should be compatible with low-end devices' limited memory resources.
- **PerfReq4**: Low power consumption. Pip-MPU's energy consumption overhead shall stay reasonable. Indeed, constrained devices are often powered on battery and the power consumption dictates their lifetime, as they are expected to operate in the wild for a long time.

## 5. PIP-MPU'S MEMORY MANAGEMENT

As detailed in Section 3, pipcore is composed of a set of services dedicated to memory management and a set of services dedicated to context switching.

Pip guarantees that the active MMU configuration respects the security requirements. This MMU configuration is collected from the metadata structures of the partitions. However, only the memory management subset changes the metadata structures during the system calls. Indeed, the context switching and the interrupt handling subsets rely on the metadata structures to set up the MMU configuration for the new active context but never modify the structures. Therefore, the latter subsets require lighter changes to set up the correct MPU configuration and the transition to the MPU-based platform is almost transparent. Hence, this section relates the transition from the MMU-based Pip to the MPU-based Pip-MPU only for the memory management subset.

### 5.1 Analogy between the nested compartmentalisation framework and Pip-MPU

The framework proposed in [10] provides design guidelines for setting up nested compartmentalisation as well as an API to call the services provided by the compartmentalisation entity. In the framework, userland components can create subdomains out of their own memory space. In this way, the analogy is direct between Pip child partitions and framework subdomains. Subdomains in the framework are created on the fly, like child partitions, which satisfies *FuncReq1*. Furthermore, the framework has been chosen because it is MPU-based without hardware modifications and as such fits perfectly COTS systems as required by *SecReq2*, *HWReq1* and *HWReq2*. In addition to that, they both claim minimality in line with *SecReq3*. For the framework, the compartmentalisation entity is specialised in providing only the minimal set of required memory isolation primitives that Pip-MPU can reuse to provide memory isolation respecting Pip's security requirements. At last, the computational complexity of the framework's services fulfils the bounded execution requirements required by *PerfReq2*.

To summarize, the framework already satisfies the requirements *FuncReq1*, *SecReq3*, *HWReq1* and *PerfReq2*. Furthermore, it partially satisfies *SecReq1* because the subdomains follow Pip's *Vertical Sharing* property and the framework also protects the privileged compartmentalisation entity and its metadata structures responsible for the MPU configuration against userland accesses, which is equivalent to Pip's *Kernel Isolation* property.

The remaining security requirements (*SecReq4* and Pip's *Horizontal Isolation* property) are covered by the framework's security policy specialisation. In a second phase, the framework implementation is guided by all the remaining requirements related to performance metrics, which are evaluated in Section 6.

## 5.2 Framework security policy specialisation

The framework needs to be specialised to fully satisfy Pip's security requirements. The specialisation occurs within the system calls and in the metadata structures.

*SecReq1* requires each child partition to be isolated from other child partitions stemming from the same ancestor partition. We must operate a framework specialisation to restrict shared memory with and between child partitions to fully embrace Pip's security policy. We decided to reflect *SecReq1* in the block sharing attributes of Pip's metadata structures. A unique block field identifies the child partition with whom the block is shared. The system calls then retrieve this single value as the only possible child partition the current partition could share this block with. Hence, from the metadata structure itself, it is impossible to share a block with multiple children, satisfying the requirement. As a consequence of the specialisation, we modify the framework's API. We removed the child partition identification for the system call retrieving shared blocks since only one child can hold a shared block.

*SecReq4* requires to restrict the access permissions updates. In the framework, access permissions are set when adding blocks to a subdomain as done with Pip, but without restrictions. In our specialisation, the read, write and execute rights can never be elevated (but can still be lowered) guarded by additional logic in the block sharing system call.

## 5.3 Implementation guidance

This section details the key design choices that oriented the framework's implementation. For the framework implementation, we opted for the user manual block protection. It consists in the development of the system call `mapMPU` which selects one of the partition's blocks to be protected by a given MPU region. `mapMPU` feeds a dedicated list present in each partition, registering all the blocks that should be enabled when the partition runs. We rejected the automated alternative proposed in the framework. It consisted in automatically reconfiguring the MPU when a memory fault occurred with another block covering the faulted address (see Section 4.C.1 in the framework paper). This is due to the limited number of MPU regions that cannot be configured to protect all the partition's blocks at the same time. In our adapted API, a memory fault is always legitimate and in such case the user is to be blamed for not having selected the correct blocks to be active in the MPU. Opting for this manual alternative increases code complexity with an additional system call. However, we expect this complexity to be negligible because we believe only a few blocks will be enabled during a partition lifetime, and more than that, it increases determinism as required by *PerfReq2*. As argued in the framework description, it also prevents us choosing a block selection algorithm that would need to be too generic.

For Pip's adaptation, we decided to keep the same nomenclature for similar objects. We already mentioned the direct transposition from protected memory spaces to partitions. By taking over the same names for equivalent metadata structures and API, Pip-MPU revives Pip's conceptual frame.

For formal verification purposes, and Pip particularly, we implemented the code directly in the Coq proof assistant. The framework's system calls settled in pipcore. As every function had to be written in Coq for later verification, it had to be adjusted to a functional environment and recursive loops. This impacts performances as well, as recursive functions use more stack memory than loops. Future works encompass a better expressiveness of the language to avoid wasting memory.

For memory purposes, we decided to combine the metadata structures, while in Pip the blocks' attributes are split into distinct structures. The rationale in Pip was to keep the MMU configuration separated from the rest of the blocks' attributes so to load the MMU directly by pointing to the new configuration. On the contrary, the MPU needs to be loaded register by register so mixing the blocks' attributes has no consequences and helps to reduce fragmentation.

For performance purposes, we enhanced the metadata structures to carry information decreasing the system calls' complexity and overall speed up the code. The first major enhancement comes by chaining blocks in a partition to their shared counterpart in a child partition. This direct link avoids going through the whole metadata structure in search for the shared block in the child partition, reducing from a O(n) complexity explained in the framework to a O(1) with the downside of adding a pointer to each block entry. This operation must be accelerated since it can be used heavily during an inter-partition communication. Moreover, in order to significantly speed up the MPU configuration, we introduced a second MPU list, besides the list registering the manual MPU mapping explained above. This second list leverages the MPU packet configuration feature, allowing a fast configuration by setting up the MPU regions fourth by fourth. It consists of a pair of register values to be slammed directly in the MPU registers, instead of configuring each MPU entry one by one by retrieving the information in the metadata structures. This second list is always updated in the system calls, at the same time as the first list. Furthermore, we limited the number of entries in the metadata structures to 64 per partition, setting the upper bound to the linear search in this structure. The linear search is needed when looking for a specific block, for example when sharing a block to check its access permission rights.

### 5.4 Pip-MPU's memory management API

The API includes the nine system calls inherited and specialised from the compartmentalisation framework, mixed with Pip's original API and naming convention: `createPartition / deletePartition`, `prepare/collect`, `addMemoryBlock/ removeMemoryBlock`, `cutMemoryBlock/mergeMemoryBlocks`, `mapMPU`. Pip-MPU differs from Pip with additional system calls that allow a partition to finely fragment its own memory space. It also differentiates from Pip by having system calls that act on the active partition itself whereas with Pip the active partition can only act on a child partition. Pip-MPU's system calls perform the necessary operations to set up isolated memory spaces by:

**Creating/deleting a child partition** A partition can create a child partition at any time. The creation occurs by designating one block of the parent partition's memory space to hold the child partition's global metadata. Hence, the number of child partitions is limited by the number of memory blocks in a memory space that is a value bounded by the framework. The global metadata, inherited from the framework, comprises: a link to the parent partition, the number of available slots to register memory blocks, the first available slot, references to its inner metadata structures that list the blocks, the number of configured inner metadata structures and the active MPU configuration. The parent partition always has prevalence over the child partition and can decide to delete (kill) a child partition at any moment. When deleting a child partition, the parent partition retrieves all the child's memory blocks.

**Preparing/retrieving the partition's inner metadata structures** Once a child partition is created, it needs the previously mentioned inner metadata structures to hold the information about its memory space. An inner metadata structure comprises the list of memory blocks in the memory space and their attributes (block location and size, access permission rights, accessibility, sharing attributes and origin). Via Pip- MPU, the parent partition can configure a memory block to become an inner metadata structure and give it to a child partition, in a similar fashion as with the global metadata structure seen above. The procedure is very similar in Pip, however, in the latter case, the inner metadata structures are subdivided into four single structures to differentiate the sharing attributes and additional optimisation metadata structures from the rest of the block attributes. This subdivision stems from the metadata structures matching the MMU page tables leveraging the MMU to accelerate information retrieval. It does not influence Pip since the MMU references millions of pages. However, it has a severe consequence for limited memory blocks in Pip-MPU and the framework advocates to merge the divided structures to save some memory blocks. Moreover, Pip-MPU stands out from Pip in these system calls since a partition can also prepare itself. This feature is fundamental to extend the list of memory blocks during runtime and to only use the memory that is strictly necessary at a given moment. This is not an issue in Pip since the MMU page tables already provide extension possibilities by filling a page table level.

**Adding/removing memory blocks to/from a child partition** Likewise Pip, a partition can share a memory block with a child partition. However, due to the lack of virtual memory, the parent partition does not know where to map a memory block in the child partition's inner metadata structures. Indeed, the list of all available slots in the child partition is dynamic and outside the control of the parent because of system calls done on itself (i.e. the child partition could have used a slot to prepare itself). Pip-MPU is in charge of the mapping at the first available slot referenced in the global metadata structure. The compartmentalisation framework anticipated this reference to the first available slot in order to avoid searching for it through the whole list of memory blocks. Pip-MPU also distinguishes from Pip from the fact that all the memory blocks cannot be enabled in the MPU at the same time. As explained in the previous section, Pip-MPU includes an additional system call so that a partition can specifically select which blocks to map in the MPU at a given moment. On the contrary, Pip does not struggle with enabled memory blocks because all mapped pages in a memory space are protected by the MMU.

**Cutting/merging back memory blocks** Pip-MPU completely differs from Pip in this last system call category. Indeed, the compartmentalisation framework features the fragmentation of a partition's inner memory space by cutting owned memory blocks. This is a direct consequence of the use of physical memory compared to virtual memory where pages are fixed-sized and always exist. In Pip-MPU, the memory blocks are crafted on the go and have a variable size down to the fine-grained resolution of an MPU region (32 bytes). Coupled with the feature to prepare metadata structures for itself, a partition can cut as many blocks it desires until reaching a maximum defined at compile-time.

## 6. EVALUATION

We evaluate our solution by implementing a Pip-MPU prototype on a device based on an ARMv7 Cortex-M processor and by comparing it to a baseline scenario without Pip. The goal of the evaluation part is to answer the following questions: 1) Is the solution usable in practice to be implemented for constrained objects? 2) What are the solution's costs and benefits in terms of performance (processor cycles, energy consumption) and system overlay (size, lines of code, initialisation time)?

### 6.1 Experimental setup

Our prototype runs on an nRF52840 DK (Nordic Semiconductor) board [23]. The board is built around an ARM Cortex-M4 CPU (ARMv7-M architecture) running as fast as 64 MHz with 1 MB of Flash and 256 kB of RAM, with an MPU composed of 8 MPU regions.

We perform static and dynamic analyses on 4 benchmark applications out of the Embench IoT benchmark suite applications [24]: ahamont64, crc32, nsichneu, primecount. We directly use the source files [25] without any modifications. They have been selected because the benchmark suite is free and open-source, the applications represent deeply embedded systems, they are compatible with our system constraints, they run on bare-metal and they don't have any output streams. They also do not use the Floating Point Unit (FPU), even if one is present on our board but our prototype does not support it yet.

The evaluation consists of two scenarios running an application 1) in Pip's root partition 2) in a child partition. The root partition sets all applications in the unprivileged userland, making it impossible for them to run privileged operations. The child partition further restricts the memory attributed to the application, with the cost of abstraction. We compared each scenario against our baseline scenario consisting in running the benchmark application in the following configuration: privileged mode, without Pip and after the same system initialisation phase. The test application is regularly interrupted by the SysTick clock every 10 ms which triggers either a void handler in the baseline scenario or Pip-MPU's interrupt management handler in the Pip scenarios. As an end result, we present the total overhead induced by the use of Pip-MPU at different abstraction level for each evaluation metric. The CPU runs at a speed of 64MHz and each benchmark application is launched successively several times within a scenario to strengthen the results disparities and extend the experiment. An experiment associates a benchmark application with a scenario. We distinguish four phases in the experiment illustrated in Figure 3: the system initialisation phase (boot), the benchmark initialisation phase (the launch of the root partition and the child partition), the test phase that is the benchmark executing for several runs, and the benchmark end phase which stops the experiment and sends the collected data to the main computer driving the evaluation. Final post-mortem analysis is carried on with all the data collected from all the experiments to extract the information and generate statistics reports.

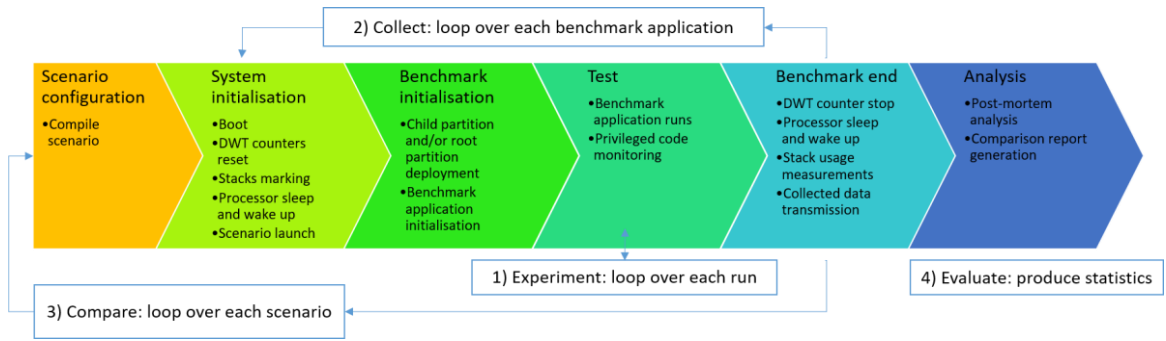

Figure 3. Evaluation phases. The evaluation consists in conducting the experiment on each benchmark application in each scenario and finally analysing all the data.

### 6.2 Evaluation results

We wrote specific Python scripts to conduct the evaluation phase, cross-mixed and adapted from the scripts and tools offered by Embench and BenchIoT [26]. In this section, we describe the monitored metrics and how we collected the data. The final results present Pip's raw overhead in Table 2 and what performance costs to expect in Table 3.

The source lines of code (SLOC) are the number of C lines of code counted after removing all comments and empty lines from the C source files by using the gcc `-fpreprocessed` option. They include lines containing only brackets, global variables and the function parameters that could spread on several lines (though remain limited). Table 2 presents the SLOC and size (in bytes) of Pip-MPU alone.

Stack usage is monitored by identifying the software components' stacks (main stack and app stack) and by marking them with a pre-defined value. As the stack is growing one address after the other, the last position where this value has been updated is the stack bottom address which witnesses the usage. In addition to the root partition's metadata structures, Pip-MPU's memory footprint also encompasses the metadata structures needed to create any runnable partition (the structures holding the list of blocks and attributes, as well as global partition data). The memory footprint is computed through formulas explained next. When the number of blocks in a partition grows by cutting or receiving memory blocks, the latter need to be registered in supplementary structures of size $S$ in bytes. Each supplementary structure can hold a constant number of blocks $C$. Hence, for a partition of $B$ blocks, we get $K + (B \bmod C) \times S\ bytes$ with K incompressible metadata. In our implementation, $K = 640$, $C = 8$, $S = 512$. As any partition requires a minimum of one metadata structure to hold the first blocks, it leads to a minimum memory footprint in RAM for each partition of 1152 B, including the root partition. Furthermore, as the number of metadata structures for a partition is bounded by $MaxMS$ at compile-time, the maximum memory footprint for a given partition is $K + MaxMS \times S$. Applied to our system, it gives a maximum footprint of $640 + 8 \times 512 = 4736$ B. More than that, $MaxMS$ also dictates the maximum number of blocks a partition can hold with the formula $C \times MaxMS$. For our system implementation, a partition can register $8 \times 8 = 64$ blocks.

Table 2. Pip-MPU raw overhead. To compute Pip-MPU's size and memory footprint, the -Os optimisation flag was used.

|  | SLOC of C | Size (B) |
|---|---|---|
| **Memory footprint in Flash** |  |  |
| pipcore (translated from Coq) | 2483 | 5804 |
| Pip handlers | 789 | 908 |
| MAL | 843 | 1996 |
| Pip init | 71 | 772 |
| Pip data + bss | - | 64 |
| Total Pip-MPU size | 4186 | 9544 |
| **Memory footprint in RAM (B)** |  |  |
| Pip-MPU stack usage | 516 |  |
| Metadata structures: |  |  |
| - Per partition | 640 + (B $mod$ 8) × 512 |  |
| - Min per partition | 1152 |  |
| - Max per partition | 4736 |  |
| **Deployment (#cycles)** |  |  |
|  | **In root** | **In child** |
| Pip-MPU initialisation | 99022 | 165582 |

For the performance metrics of Table 3, we run the benchmark application configured for each scenario (baseline, in root partition and in child partition). Each time we execute 3 runs in a row within the same experiment to collect data during at least 20 seconds (each benchmark application executes during 5-7 seconds). We launch each experiment 5 times and perform statistics on the results (average $\mu$ and standard deviation $\sigma$). The indicated overhead is the observed average overhead computed for each scenario compared to the baseline, e.g. the average on all benchmark applications of the average overhead on all runs.

The cycles count are retrieved from the Data Watchpoint and Trace (DWT) unit of the processor. We initialise the count just before the launch of the benchmark application and collect its value after the end of the initialisation phase and when the application is finished. The end of the initialisation phase marks the test phase, from where the benchmark application is executing. For the baseline scenario, the initialisation phase is almost void since it just calls the benchmark application. Moreover, the baseline scenario is always executing in privileged mode so the cycles count is fully privileged. On the contrary, in the Pip scenarios, the privileged cycles are monitored by counting the cycles only spent in Pip-MPU. We provide the ratio of privileged cycles over the total cycles from i) Pip-MPU's start and ii) only during the test phase. They are compared to the entirely privileged baseline.

Table 3. Performances comparison (versus baseline). The test application is either executed in the root partition or in the child partition, compared to the baseline.

| Metrics | In root | In child |
|---|---|---|
| **Cycles** | | |
| Cycles overhead: | | |
| i) in total | $\mu = 76302131$ | $\mu = 74538344$ |
| | $\sigma = 67494444$ | $\sigma = 73634323$ |
| | (+16.31%) | (+16.4% ) |
| ii) during test | $\mu = 76203107$ | $\mu = 74372762$ |
| | $\sigma = 67495112$ | $\sigma = 73634647$ |
| | (+16.29%) | (+16.36%) |
| Privileged cycles over total cycles ratio: | | |
| i) in total | $\mu = 0.86\%$ | $\mu = 0.92\%$ |
| : | $\sigma = 3.8 \times 10^{-5}\%$ | $\sigma = 3.3 \times 10^{-5}\%$ |
| | (-99.14%) | (-99.08%) |
| ii) during test | $\mu = 0.87\%$ | $\mu = 0.92\%$ |
| | $\sigma = 3.9 \times 10^{-5}\%$ | $\sigma = 3.3 \times 10^{-5}\%$ |
| | (-99.13%) | (-99.08%) |
| **Energy consumption during test** | | |
| Total energy overhead | $\mu = 24.76 mJ$ | $\mu = 26.6 mJ$ |
| | $\sigma = 22.42 mJ$ | $\sigma = 23.00 mJ$ |
| | (+16.7%) | (+18.4%) |
| Energy overhead due to MPU | $\mu = 0.05 mJ$ | $\mu = 0.07 mJ$ |
| | $\sigma = 0.16 mJ$ | $\sigma = 0.11 mJ$ |
| | (+0.03%) | (+0.04%) |
| **Security** | | |
| Accessible application memory over total memory ratio: | | |
| - Flash (code) | 99.0% | 6.27% |
| | (-1.0%) | (-93.73%) |
| - RAM (data) | 99.35% | 1.9% |
| | (-0.65%) | (-98.1%) |

The accessible memory areas represent the memory a partition has access to. The application in the privileged baseline has access to the whole memory whereas by using Pip-MPU the accessible memory areas are the blocks of the memory space. For the root partition, the accessible memory includes the whole memory minus the TCB (Pip-MPU and boot components). From there on, the root partition, as any other parent partition, decides which memory blocks to pass on to its children, thereby controlling their accessible memory areas.

The energy consumption has been monitored using the Power Profiler Kit I (PPKI) [27] mounted on the nRF52840 DK board (Figure 4). The PPK provides current measurements at 77 kHz with 4 measures average that we multiply with a fixed voltage and integrate over time to get the total energy consumption. As the benchmarks use semihosting to send the performance data (cycles and stack usage) to the computer for analysis, the debugger remains active. However, no input or output is performed during the test phase. Furthermore, our set-up includes an additional nRF52840 DK board to interface with the PPK which sends the measurements to the computer. We used a PPK library [28] to trigger the measurements because the desktop application was not stable enough for our experiments and it eased the integration with our python scripts. Nevertheless, as an upgraded version of the PPK (PPKII) was released some years ago, the library is not maintained anymore and the integration required to find a good match between the

PPK's firmware version and the library and its dependencies. For our analysis, the energy consumption is solely measured during the test phase. We mark this phase by setting the processor in deep sleep mode before and after the test phase and wake it up with an external timer. In this way, we can easily identify the test phase from the current measurements with significant current drops during the sleep phases (around 6mA during the test phase down to $\mu$A when sleeping).

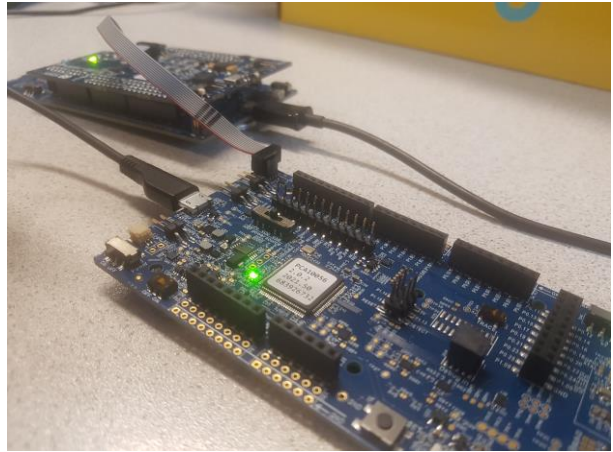

Figure 4. Test bed. In the foreground, the nRF52840-DK controlling the PPK. In the background, the nRF52840-DK executing the test application on which is mounted the PPK.

### 6.3 Discussions and limitations

The figures presented in the previous section are valuable information to consider a port on Pip-MPU.

Pip-MPU takes respectively 1664 B (data, stack, root partition metadata structures) and 9544 B (code) of the available 256 kB RAM and 1 MB of Flash. It then fits easily the constraints of our targets (around 3.3% RAM and 3.8% Flash of Class 2 IETF devices) and leaves enough space for more complex applications, thereby fulfilling *PerfReq3*. Pip-MPU is smaller than PISTIS or TockOS and comparable to the smallest OS kernels with a size of around 6 kB for pipcore. Pip-MPU's minimality, required by *SecReq3*, is therefore satisfied. Hence, we expect a good ratio for Pip-MPU's size relative to the size of rich OSs and their applications ported on Pip-MPU. To be noted, we considered scenarios with a correct test application, without triggering faults or using the partial MPU reconfiguration feature inherited from the nested compartmentalisation framework. We expect a stack usage increase in such cases.

The accessible memory areas metric shows the extent of the attack surface. In the baseline scenario, since the application is privileged, it can access 100% of the memory. On the contrary, when using Pip-MPU, the partition becomes unprivileged and is limited by the MPU. For the root partition, this value decreases by about 1%. Indeed, the root partition owns the whole memory except the parts reserved for Pip-MPU. The further away the active partition is from the root partition, the more the parent partition can restrict the accessible memory and better is this metric. For the child partition in our implementation, we reduced its accessible memory area to respectively 2% and 6% of the RAM and Flash areas. This means this child partition loses more than 94-98% of the memory that was accessible in the privileged baseline scenario.

The evaluation reveals a minimum memory footprint in a parent partition for each new child partition of around 1 kB for our implementation. This minimum should be increased by the requisitioned entries in the parent partition to register the child's metadata structures. The additional entries may not fit in the fixed-size metadata structure holding the block attributes, leading to the creation of a new metadata structure in the parent to host these entries (supplementary 512 B in our implementation).

Pip-MPU's raw overhead is declined in two stages: the initialisation phase (for the root and child partitions) and the test phase (the running application). The initialisation phase shows an averaged initialisation phase lasting 99022 cycles ($1.5ms@64MHz$) and 165582 ($2.6ms@64MHz$) respectively for the root partition and the child partition. This represents the pure overhead of Pip-MPU's initialisation time over the baseline, resonating with *PerfReq1*. Furthermore, we observed an execution overhead for the test phase of about 16 % caused by Pip-MPU's restoration context sequence when receiving the SysTick interrupt. This latter value should be appreciated within the tested scenario and values are expected to be higher for a rich OS ported on Pip-MPU because of multiple interrupts causes. While the performances proved sufficient in the evaluation, there are potential improvements areas to further optimise the system calls if deemed necessary in the future by adding optimisation metadata structures similar to Pip (MMU). In addition to that, Pip-MPU forces the benchmark application to run in unprivileged mode. We observe a drop of more than 99% of the privileged cycles when using Pip-MPU that correspond to Pip-MPU's execution. The opportunities to exploit the privileged operation mode reduce as much.

Energy consumption resulted in a 17-18% increase when using Pip-MPU. Moreover, we launched the benchmarks while switching off the MPU. It showed a consumption decrease of 0.02-0.2% depending on the scenarios. It indicates that the MPU use (due to the context switching and permanent protection) does not impact significantly the power consumption. These measurements are important for IoT devices that may operate in areas without power line access and thus depend on a limited power battery. They satisfy the final requirement *PerfReq4*.

Other metrics are proposed in BenchIoT but are not evaluated here for the following reasons. First, we did not evaluate the number of sleep cycles as Pip-MPU never puts the CPU into sleep. Second, we did not include Data Execution Prevention (DEP) or the enforcement of the $W\hat{}X$ security principle, because Pip-MPU does not set them up. Indeed, the existence of such or additional security principles (like deciding which memory blocks to isolate) are strict partition design choices. Third, ROP gadgets and indirect calls are known techniques for an attacker to take control of the control flow and perform impactful attacks [29]. We evaluated the ROP gadgets and indirect calls overhead respectively to 1780 and 9 due to Pip-MPU (directly using BenchIoT's tools based on [30]). However, we do not recognise them as relevant for Pip-MPU. The rationale is that Pip-MPU's or ancestor partitions' code and data are private and invisible from the point of view of the active partition. Illegal access trials by crafted ROP gadgets end up in MPU memory faults caught by the ancestors. Furthermore, pipcore being developed in Coq before C translation, it holds characteristics of a functional programming language like high stack usage and many functions degrading these particular metrics. Hence, they do not represent for us relevant metrics. It should be noted though that Pip does not prevent ROP attacks within the partition but against Pip and the partition's ancestors. Fourth, we did not single out privileged cycles and SVC cycles as they represent the same thing for Pip. Indeed, Pip's entry points are the SVC and are the only privileged code that can run after the initialisation phase.

As a result, the preliminary analysis and the evaluation showed full compliance to Pip's requirements and those expected for resource-constrained devices. Impactful security measures like privilege segregation of user and kernel/sensitive code are sometimes not used to lower production costs or reduce energy consumption. We showed simple applications such as those used in our evaluation can directly benefit from Pip-MPU's protection with almost no effort.

The scenarios explored in the benchmarks have a maximum of one isolation level. This is sufficient for bare-metal applications but we expect another level when porting an OS. A supplementary level implies additional abstraction to go through the partition tree that might degrade the performances.

Pip-MPU entails the presence of an MPU which is a strong limitation for embedded systems without MPU. However, previous works [31] showed the MPU is present most of the time in Cortex- M3/4/7-based micro-controllers, thereby supporting the applicability of Pip-MPU. In addition to that, the compartmentalisation framework is generic to systems supporting privileged mode segregation and have an equivalent unit to the MPU. We believe our approach is then reproducible on processors from other vendors providing equivalent features.

## 7. CONCLUSION

In this paper, we present Pip-MPU, the Pip kernel variant based on the Memory Protection Unit (MPU) which does not require any hardware modification on Commercial Off-The-Shelf (COTS) systems. We achieve transposing the memory isolation offered by the MMU into MPU-based memory isolation by specializing the framework provided by Dejon et al. so that it satisfies the security requirements of Pip. We also defined and verified additional requirements which are specific to the context of constrained devices. We present our implementation which is also portable to other ARM architectures such as the ARMv8 Cortex-M architecture. Our evaluation is performed on a fully implemented prototype based on ARMv7 Cortex-M. We show that Pip-MPU reduces the attack surface from 100% down to 2% while requiring 10 KB of Flash, 550B of RAM and an overhead of 16% on both performance and energy consumption. To our knowledge, Pip-MPU is therefore the first and smallest isolation kernel for resource-constrained devices which provides nested compartmentalisation.

Currently, Pip-MPU is under formal verification by building on Pip's proof methodology. In future works, we will explore how Pip's flexibility can be leveraged to create a secure-by-design architecture for containers on low-end devices, as described in [32]. This use case differs from the typical use case for low-end devices which consists in isolating multiple code components within a single-thread and multi-tasking bare-metal application because it involves multiple parties and requires reconfiguring the memory partition during the device's lifetime. We will also explore how the isolation guarantees provided by Pip can be propagated in remote attestations for example.

## ACKNOWLEDGEMENTS

The research leading to these results partly received funding from the MESRI-BMBF German-French cybersecurity program under grant agreements no ANR-20-CYAL-0005 and 16KIS1395K. The paper reflects only the authors' views. MESRI and BMBF are not responsible for any use that may be made of the information it contains. This work was also supported by IRCICA, USR-3380 (Lille, France).

## AUTHORS

**Nicolas Dejon** is a third-year Ph.D student at the University of Lille (CRIStAL laboratory, 2XS team) in France and works full-time for Orange Labs in Caen (France) as joint collaboration (French CIFRE fellowship). He graduated in Computer Science and Engineering, specialised in Embedded and Real-Time Computer Systems, at the University of Technology of Compiègne (UTC, France) in 2018. He expanded his studies by obtaining the European Master degree EMECIS delivering a double joint title of Master in Complex Systems Engineering from the UTC and Laurea Magistrale in Ingegneria Informatica from the University of Genoa (UNIGE, Italy) in 2019.

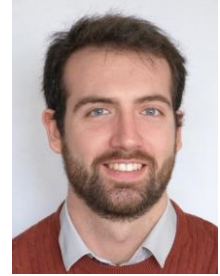

**Chrystel Gaber** received her PhD from University of Caen in 2013 in Computing Systems. After an experience as project coordinator and R&D engineer in Fime, she joined Orange as a researcher & project coordinator. She contributes to several projects related to cyber-physical security, IoT device management and certification. She participated in the FP7 project MASSIF and ensured the coordination lead of the CELTIC-PLUS project ODSI. She represented Orange in GSMA work groups related to the certification of integrated SIMs and the accreditation of SIM production sites. Currently, she contributes to the H2020 European project INSPIRE-5GPlus and leads the franco-german project TinyPART.

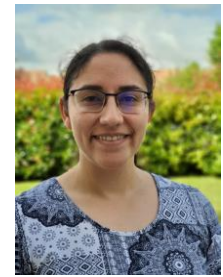

**Gilles Grimaud** is full professor at the University of Lille (France) since 2009. Before that, he obtained his Ph.D from the University of Lille 1 (France) in December 2000, awarded by the french chapter of ACM-SIGOps. He also spent 18 months (2000-2002) in the Gemplus Research Labs where he worked on next generation smardcard operating systems. In 2002, he joined RD2P/LIFL/CNRS and POPS/INRIA-Futur Research at the University of Lille as an associate Professor. He was elected vice-president of the french chapter of ACM-SIGOps (2005-2008). He is currently the scientific director of the 2XS (Extra Small Extra Safe) team (CRIStAL laboratory, Lille) since July 2012. His research interests include efficiency, safety and security of embedded systems and dedicated operating systems. In particular, his current project is related to an open source operating system that is formally proved and called PIP. Formal proof include isolation (using MMU or MPU hardware on embedded systems) and RT scheduling properties. Hardware targets include intel and Arm processor families. Past projects include the Smart and Mobile Embedded Web Server : Smews, the Camille OS and the JITS (i.e. Java In The Small) platform.

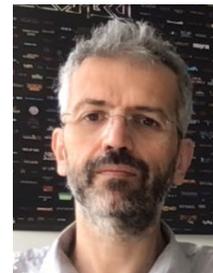